# Birch's law at elevated temperatures

Umesh C. Roy[1] and Subir K. Sarkar[1,a]

1. School of Physical Sciences, Jawaharlal Nehru University, New Delhi 110067 INDIA

a) Author to whom correspondence should be addressed. Electronic mail: ssarkar@mail.jnu.ac.in

## ABSTRACT

*Birch's law in high pressure physics postulates a linear relationship between elastic wave speed and density* and one of its most well known applications is in investigations into the composition of the inner core of the Earth using the Preliminary Reference Earth Model as the primary source of constraints. However, it has never been subjected to high precision tests even at moderately elevated temperatures. Here we carry out such a test by making use of the Density Functional Theory of electronic structure calculation and the Density Functional Perturbation Theory of calculating the phonon dispersion relation. We show that a recently proposed modification to the Birch's law is consistently satisfied more accurately than its original version. *This modified version states that it is the product of elastic wave speed and one-third power of density that should be a linear function of density*. We have studied the cases of platinum, palladium, molybdenum and rhodium with cubic unit cell and iron with hexagonal-close-packed unit cell with temperatures up to 1500K and pressures up to about 360 GPa. We also examine the genericity of the validity of *a recently proposed extension of the Birch's law according to which elastic wave speed is a linear function of temperature at a given density*. Within the error bars of our calculation, we find that this is consistent with our data for the four cubic materials at temperatures up to 3300 K.

# I. INTRODUCTION

Despite decades of research in the field the chemical composition of the inner core of the Earth is not understood with clarity [1-4]. The possible answers are strongly constrained by the Preliminary Reference Earth Model [5] which provides the values of density, pressure and the speeds of the two types of seismic body waves (P-waves and S-waves) at various distances from the centre of the Earth. The dependence of elastic wave speed on density throughout the inner core region can be extracted from this model. However, since the thermodynamic conditions prevailing in the inner core region cannot presently be reproduced in the laboratory, any proposal regarding the chemical composition of this region cannot be tested directly in the laboratory. What is usually done is to measure the dependence of elastic wave speed on density in the region of relatively low pressure and then to extrapolate it to the higher densities of the inner core region by using a law of high pressure physics called the Birch's law [6-8]. For the purpose of the present work this law may be stated as follows: (i) The dependence of the elastic wave speed on the thermodynamic conditions is entirely through density and (ii) this dependence on density is linear. While the elastic wave speed does depend primarily on density there is a weaker dependence on temperature also.

Birch's law has been tested extensively [9-28]. While it seems to hold fairly well these tests do not have the level of precision that is required to use it for the purpose of conclusive discrimination among the various candidates for the inner core composition (which usually differ by no more than five percent in their predictions of density and elastic wave speed under the inner core conditions). The latter requires both precise data from the laboratory experiments and a precise formula for extrapolation of the results to higher densities. In this work we examine the second aspect and for this we need precise enough data for a large range of densities. The range of materials tested should be as wide as possible so that we can have confidence in the genericity of the law. Presently only numerical calculations based on the Density Functional Theory (DFT) can provide such data. Recently [29] we reported on such a computational study in which it was shown that the DFT-based data had a superior level of agreement with a *modified version of the Birch's law according to which it is not the elastic wave speed alone but its product with one-third power of density that should be a linear function of density* [30]. However, our previous DFT-based calculations were carried out at zero temperature only. In applications such as inner core studies, of course, temperatures in excess of 5000K are involved. This target is not immediately achievable in laboratory experiments where the highest temperature at which these tests have been done is about 3000K [2], as far as we are aware. Outcomes of these laboratory experiments serve as valuable points of reference for parallel tests based on computation. The latter should be able to first broadly replicate these laboratory results and then improve upon them in terms of the level of precision so that perhaps a more precise description of the phenomenology can be developed. It is in this spirit that the present work has been carried out. *Here we investigate, again through DFT-based calculations, whether the superiority of the modified version of the Birch's law continues to hold good in the domain of moderately elevated temperatures*. A critical element of such computational work is having enough confidence in the methodology employed and in the present problem temperature is the key parameter that controls this aspect. As will be seen later this constrains us to keep the value of temperature below 1500K – although it is possible that this is too conservative.

A finite value of the temperature immediately makes the DFT-based calculations substantially alter their nature and makes them vastly more computationally expensive [31-50]. Hence we have limited our studies to five materials -- the primary criterion for their choice being thermodynamic phase stability with respect to crystal structure and magnetic properties throughout the range of pressures (generally between 20 to 360 GPa, but between 50 and 300 GPa for hcp-iron) and temperatures (300K to 1500K) studied [41,51-59]. Of these, four (palladium, rhodium, platinum and molybdenum) have cubic unit cell and one (non-magnetic iron) has a hexagonal-close-packed unit cell. Since we treat the lattice vibration within the framework of the quasi-harmonic approximation temperature is mostly (with some exceptions to be discussed later) kept below 1500 K to ensure the physical justifiability of the computational methodology i.e. the amplitude of lattice vibration should be sufficiently small.

The primary results of our study are the following: *After defining a suitable metric we show that , at least up 1500K, the modified Birch's law is consistently more accurate than the original version – even though the latter is satisfied fairly well.* However, the extent of improvement depends on the material and seems to show signs of gradual erosion with increasing temperature. We also investigate if *a recently proposed extension [2] of the Birch's law, according to which the elastic wave speed is a linear function of temperature at a given density,* is generically valid. We perform these tests for the four cubic materials mentioned earlier at temperatures up to 3300 K . *Our finding is that it is indeed consistent with our data in all cases if the computational error bars are taken into account.* This paper is organized as follows: In section II we present the key elements of the methodology of computation. It should be noted that there are some important points of departure from existing works. Section III presents the results while section IV contains a discussion and some critical comments .

## II. METHODOLOGY

To test the validity of the original Birch's law (or any modified version) for a given material and at a given temperature we need to calculate the speed of elastic waves for a selected set of densities [60-67]. In our case these densities are chosen so that the pressure varies approximately in the range of 20 GPa to 360 GPa for materials other than hcp-iron (for iron the range is from 50 to 300 GPa). The single crystal elastic constants, from which elastic wave speeds are eventually calculated, are obtained for both isothermal and adiabatic conditions – to be denoted by the superscripts 'T' and 'S', respectively. The reference state (**X**) at a given density and temperature, for which the elastic constants are to be calculated, is pre-stressed in general [61]. To calculate the isothermal elastic constants we use the defining equation in terms of the Helmholtz free energy (F ) per unit cell. Let $\sigma_{ij}$ and $\eta_{ij}$ denote the components of the stress tensor (in the reference state **X**) and the Lagrangian strain parameter, respectively. Then the expression of the free energy per unit cell at the temperature T, correct up to the second power of strain, is given by [61]

$$F(\mathbf{X},\boldsymbol{\eta},T) = F(\mathbf{X},\mathbf{0},T) + v_0 \left[\sigma_{ij}\,\eta_{ij} + \frac{1}{2}\,C^T_{ijkl}\,\eta_{ij}\,\eta_{kl}\right] \quad (1)$$

where $v_0$ is the volume per unit cell in the reference state **X** and $C^T_{ijkl}$ denotes an isothermal elastic constant. Summation over repeated indices is implied in the eqn.(1). To calculate the stress tensor and the isothermal elastic constants appearing in the eqn.(1) the free energy has to be computed first for various strengths of suitable kinds of distortion. This is followed by a best fit procedure to extract the values of these parameters.

Table I: Expression for the free energy/unit cell for different types of distortion.

| System | ElaStic software notation | $\eta_1$ | $\eta_2$ | $\eta_3$ | $\eta_4$ | $\eta_5$ | $\eta_6$ | $[F(\mathbf{X},\boldsymbol{\eta},T) - F(\mathbf{X},\mathbf{0},T)]/v_0$ |
|---|---|---|---|---|---|---|---|---|
| Cubic | $\eta^{(1)}$ | θ | θ | θ | 0 | 0 | 0 | $-3p\theta + \frac{3}{2}(C_{11}+2C_{12})\theta^2$ |
|  | $\eta^{(8)}$ | θ | θ | 0 | 0 | 0 | 0 | $-2p\theta + (C_{11}+C_{12})\theta^2$ |
|  | $\eta^{(23)}$ | 0 | 0 | 0 | 2θ | 2θ | 2θ | $6C_{44}\theta^2$ |
| Hcp | $\eta^{(1)}$ | θ | θ | θ | 0 | 0 | 0 | $(2\tau^{xy}+\tau^z)\theta + \frac{1}{2}(2C_{11}+2C_{12}+4C_{13}+C_{33})\theta^2$ |
|  | $\eta^{(3)}$ | 0 | θ | 0 | 0 | 0 | 0 | $\tau^{xy}\theta + \frac{1}{2}C_{11}\theta^2$ |
|  | $\eta^{(4)}$ | 0 | 0 | θ | 0 | 0 | 0 | $\tau^z\theta + \frac{1}{2}C_{33}\theta^2$ |
|  | $\eta^{(17)}$ | 0 | 0 | θ | 2θ | 0 | 0 | $\tau^z\theta + \frac{1}{2}(C_{33}+4C_{44})\theta^2$ |
|  | $\eta^{(26)}$ | θ/2 | θ/2 | -θ | 0 | 0 | 0 | $(\tau^{xy}-\tau^z)\theta + \frac{1}{2}(\frac{C_{11}}{2}+\frac{C_{12}}{2}-2C_{13}+C_{33})\theta^2$ |

The choice of the reference state (**X**) for a given density is a nontrivial task in the case of hcp-iron. Here density specifies only the combination $a^2c$ of the two lattice parameters a and c. We choose, for a given density, that value of a (or equivalently c) which minimizes the Helmholtz free energy per unit cell (making sure that the free energy is computed only after the positions of the atoms in the unit cell are optimized). It can be demonstrated that this corresponds to a minimization of the Gibbs free energy per unit cell for a situation when the stress tensor is hydrostatic.

Exploiting the symmetries of the stress tensor, the strain parameter and the elastic constants with respect to the exchange of indices eqn. (1) can be rewritten, in the Voigt notation, as

$$F(\mathbf{X},\eta,T) = F(\mathbf{X},\mathbf{0}, T) + v_0 [\sigma_a \eta_a + \frac{1}{2} C^T_{ab} \eta_a \eta_b] \quad (2)$$

where summation over repeated indices is implied. We take the independent elastic constants for the cubic crystal and the hcp crystal to be {$C_{11}$, $C_{12}$, $C_{44}$} and {$C_{11}$, $C_{12}$, $C_{13}$, $C_{33}$, $C_{44}$}, respectively. The set {$\eta_1$, $\eta_2$, $\eta_3$, $\eta_4$, $\eta_5$, $\eta_6$} is chosen (table I) according to the prescriptions of the ElaStic software package [66]. In the eqn.(1) the stress tensor $\sigma_{ij}$ = -p $\delta_{ij}$ for the cubic case whereas, for the hcp case, the stress tensor is diagonal with $\sigma_{11}$ = $\sigma_{22}$ = $\tau^{xy}$ and $\sigma_{33}$ = $\tau^z$ (in table I). Due to the particular way we optimize the hcp unit cell for iron in our calculations the calculated stress tensor should be hydrostatic and $\tau^{xy}$ should be exactly equal to $\tau^z$. In actual calculations they are slightly different due to the finite precision of the numerics. In fact the difference between the values of $\tau^{xy}$ and $\tau^z$ is a good indicator of the quality of the calculation of a and c for the equilibrium unit cell for a given density and temperature in the case of hcp-iron (see the Supplementary Material). Inspection of the expressions of the free energy in the presence of distortion (see table I) also shows that p or $\tau^{xy}$ or $\tau^z$ appears in the expressions for more than one type of distortion. Since the evaluation of the parameters appearing for the different types of distortion are done independently, this redundancy provides crucial checks for the accuracy and reliability of our calculations. We choose K uniformly spaced values of θ (defined in table I) from -θ$_{max}$ to + θ$_{max}$ -- with the values of {K,θ$_{max}$} being {11, 0.0107} for Pd and Rh and {9, 0.01} for Mo, Pt and hcp-Fe. For each type of distortion a cubic polynomial is fitted to the data for the free energy vs. θ (no linear term is included in the fitting function when it is supposed to be absent (according to table I)) and from its coefficients information regarding the components of the stress tensor and the elastic coefficient tensor is extracted.

The free energy per unit cell is a sum of several terms. The general form is

$$F(\mathbf{X},\eta,T) = E_0 + E_{el}(T) - T S_{el} + F_{ph} \qquad (3)$$

Here $E_0$ is the electronic ground state energy per unit cell (when more than one atom is present in the unit cell, this is the electronic ground state energy minimized with respect to the atomic positions). $E_{el}(T)$ is the excess electronic energy due to excitations at finite temperature and is defined as

$$E_{el}(T) = <E_e(T)> - <E_e(0)> \qquad (4)$$

where

$$<E_e(T)> = \int \varepsilon n(\varepsilon) f(\varepsilon,T) d\varepsilon . \qquad (5)$$

Here $f(\varepsilon,T)$ is the Fermi occupation function $[\exp((\varepsilon-\mu)/k_B T)+1]^{-1}$ and $n(\varepsilon)$ is the electronic density of states per unit cell at the energy $\varepsilon$. The chemical potential $\mu$ at the temperature T is obtained numerically by ensuring that the integral $\int n(\varepsilon) f(\varepsilon,T) d\varepsilon$ equals the total number of valence electrons per unit cell. The value of $<E_e(0)>$ is obtained by the graphical extrapolation of $<E_e(T)>$, calculated at finite values of T, to T = 0. The expression for the electronic entropy $S_{el}$ in the eqn.(3) is

$$S_{el} = -k_B \int n(\varepsilon) [f(\varepsilon,T) \ln(f(\varepsilon,T)) + (1-f(\varepsilon,T)) \ln(1-f(\varepsilon,T))] d\varepsilon \qquad (6)$$

The phonon free energy (including zero point energy) $F_{ph}$ in eqn.(3) is given by the expression

$$F_{ph} = k_B T \int g(\omega) \ln [2 \sinh(h\omega/4\pi k_B T)] d\omega \qquad (7)$$

, where $g(\omega)$ is the phonon density of states at the (circular) frequency $\omega$ and is normalized so that $\int g(\omega) d\omega$ is equal to three times the number of atoms per unit cell. The key physical assumption behind writing the eqn.(7) is what is normally referred to as the quasi-harmonic approximation i.e. the amplitude of atomic displacements is assumed to be small enough to permit a quantum statistical description of the vibration of the lattice in terms of a gas of non-interacting phonons.

Finally, the adiabatic elastic constants are related to the isothermal ones via the formula [61]

$$C^S_{ijkl} = C^T_{ijkl} + (Tv_0/C_v) \, v_{ij} \, v_{kl} \qquad (8)$$

where

$$v_{ij} = (\partial \sigma_{ij}/\partial T)_{\{\eta\}=0} \qquad (9)$$

and $C_v$ is the specific heat per unit cell under the condition of constant volume. The derivatives $v_{ij}$ are calculated numerically after calculating the stress tensor at an equally spaced set of temperatures situated symmetrically around the target temperature for which the derivatives are needed. The specific heat per unit cell $C_v$ has contributions from electronic and phononic excitations. Thus $C_v = C_{v,el} + C_{v,ph}$ where

$$C_{v,ph} = k_B \int (h\omega/2\pi k_B T)^2 \, g(\omega) \, \chi(\omega,T) \, (1+ \chi(\omega,T)) \, d\omega \qquad (10)$$

, with $\chi(\omega,T) = [\exp(h\omega/2\pi k_B T)-1]^{-1}$. The electronic contribution $C_{v,el}$ is given by the expression

$$C_{v,el} = \int [(1/k_B T)(d\mu/dT) + (1/k_B T^2)(\varepsilon-\mu)] \varepsilon \, n(\varepsilon) \, f(\varepsilon,T) \, (1- f(\varepsilon,T)) \, d\varepsilon \qquad (11)$$

where $f(\varepsilon,T)$, as before, is the Fermi occupancy factor. The temperature derivative $(d\mu/dT)$, appearing in the eqn.(11), is calculated by numerically evaluating its expression $-(1/T) [\int(\varepsilon-\mu)n(\varepsilon) f(\varepsilon,T) (1- f(\varepsilon,T))d\varepsilon / \int n(\varepsilon) f(\varepsilon,T)( 1- f(\varepsilon,T))d\varepsilon]$.

All calculations involving the application of the DFT are done by making use of the Quantum Espresso package [65]. This includes the computation of the vibrational spectrum using the Density Functional Perturbation Theory [67]. The distorted unit cells required for the calculation of the elastic constants are generated using the ElaStic software [66]. However, actual calculation of free energy and all subsequent analysis of data are done manually by us.

The targeted level of convergence in the calculation of free energy is 1 meV. Tests were performed on the convergence with respect to the **k**-grid and the **q**-grid for the electronic and the vibrational spectrum. The final values used are shown in the Supplementary Material. In the case of iron there are two atoms per unit cell. In this case optimization of the atomic positions also has to be done. The calculation of the P-wave speed in polycrystalline aggregates from the values of single-crystal elastic constants is explained in the Supplementary Material.

## III. RESULTS

We have studied five elemental solids – of which four (palladium, platinum, rhodium and molybdenum) are cubic and one (non-magnetic iron) is hexagonal-close-packed. To ensure that the quasi-harmonic approximation is uniformly accurate at all densities we have limited the temperature to 1500K -- although it is quite likely that, at least at the highest densities that we use, the approximation is quite accurate even at much higher temperatures. As mentioned earlier the isothermal elastic moduli are calculated from the coefficients of the best fit polynomials for the variation of the Helmholtz free energy as a function of the strength of distortion. In general these polynomials are taken to be cubic. However, when the linear term is expected to be absent (see table I) it is omitted from the fitting function also. An example of such fits for the cubic case is shown in the Supplementary Material.

All the isothermal elastic moduli and other information required to calculate the adiabatic elastic moduli are available in a tabular form in the Supplementary Material. From these we first calculate the adiabatic generalized Birch coefficients using the procedure described in the Supplementary Material. In the literature it is common to fit the computed free energies to some general formulas [68-69]. We do not do it. This frees us from reliance on the accuracy of any such formulas but it also places demand for high accuracy in the free energy data and ample caution in the calculation of various derivatives. In those cases where our calculations have some overlap with existing literature on computation and/or measurement of elastic constants, especially at finite temperatures, we have performed consistency checks with a sampling of such works. This includes data on elastic constants: (i) in figure 12 of ref.36 for hcp-iron, (ii) in figure 4 of ref.41 and tables I and III of ref.70 for palladium, (iii) in table II of ref.34, tables I and III of ref.70 and fig.3 of ref.71 for platinum, (iv) in table I of ref.72 and tables I and III of ref.70 for rhodium and (v) in figures 4 and 11 of ref.33, tables I and III of ref.70, tables I and II of ref.73 for molybdenum. In these comparisons the extent of disagreement with our calculations of the elastic constants rarely exceeds ten percent and often is much less (see the Supplementary Material).

Having performed these checks we proceeded to the calculation of the direction-averaged longitudinal acoustic (LA) speed in a single crystal ($<V_{LA}>$) and the P-wave speed ($V_P$) for elastic waves in a polycrystalline medium. These two types of speed are quite close to each other in numerical value but are not quite the same conceptually and are of relevance to different kinds of experiments. The first one ($<V_{LA}>$), crudely speaking, corresponds to the average value of the longitudinal acoustic speed that might be measured in an inelastic scattering experiment using a poly-crystalline sample with truly random grain alignment (see [74] for a more careful analysis) . The second one ($V_P$) describes the speed of seismic wave propagation in an extended medium like the Earth. We subject these two kinds of speed to independent analysis.

Plots of $<V_{LA}>$ and $V_P$ against density are shown for the five materials studied in figures 1 through 5. It may be noted that there are no error bars associated with the density co-ordinates since these values are pre-selected and are not results of any calculation. The error bars for the speed co-ordinates are always comparable to or smaller than the sizes of the symbols representing the data points. Hence they are not shown in the plots [The methodology used in the calculation of the error bars is explained in the Supplementary Material]. For iron our calculations were done only at 300K and 1500K. For the cubic materials the temperatures used were 300K, 900K and 1500K. However, for the sake of visual clarity, only the data at 300K and 1500K are presented in the figures for all the five cases.

The first point to note is that (within the range of densities studied) , for a given density, the elastic wave speed can both increase or decrease with temperature. It depends on the material. For iron it is known from several previous studies, both experimental and computational, that the speed goes down with increasing temperature (see ref.2, for example and the references therein). Our results are qualitatively consistent with these known facts – although the extent of reduction in our calculations is somewhat less than those reported in the experiments. Out of the remaining four cases, speed goes down with increasing temperature for rhodium whereas it goes up for platinum and palladium. For molybdenum there is a transition between these two kinds of behaviour just below the highest density we have studied.

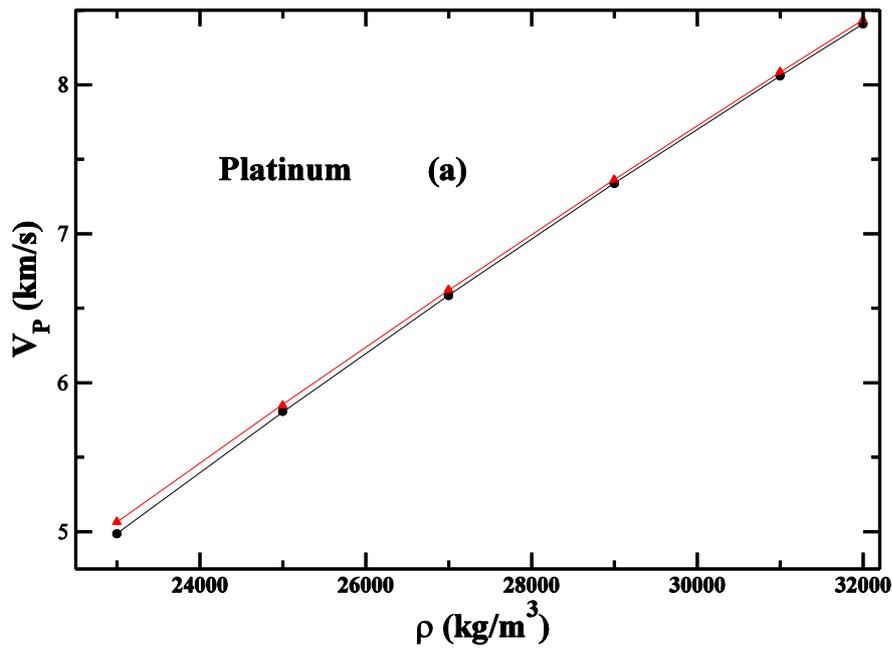

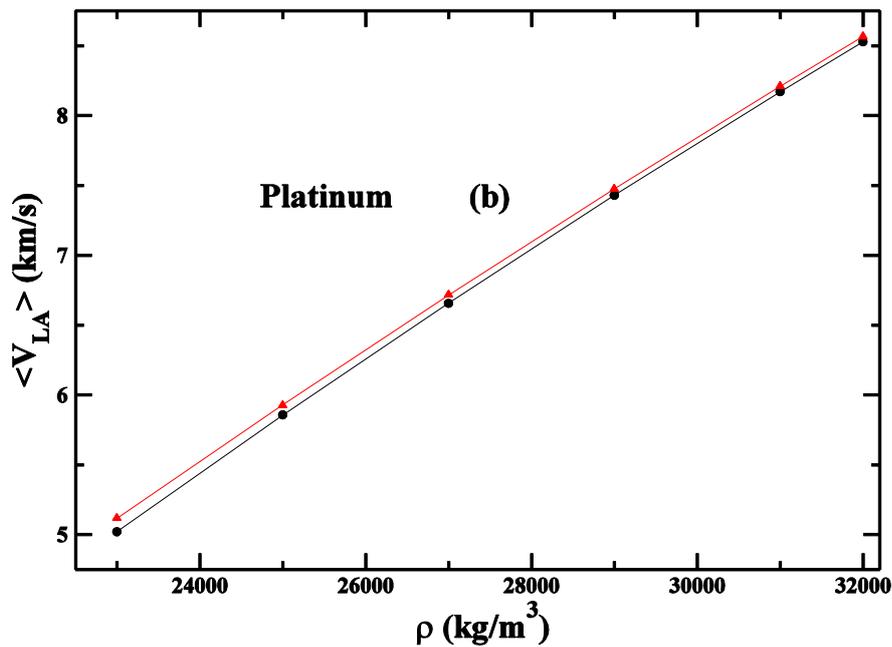

Fig.1: [Color online] (a) P-wave speed ($V_P$) and (b) direction-averaged longitudinal acoustic speed ($\langle V_{LA} \rangle$) against density ($\rho$) for platinum at 300 K (filled circle) and 1500K (filled triangle). Continuous lines are guides for the eye.

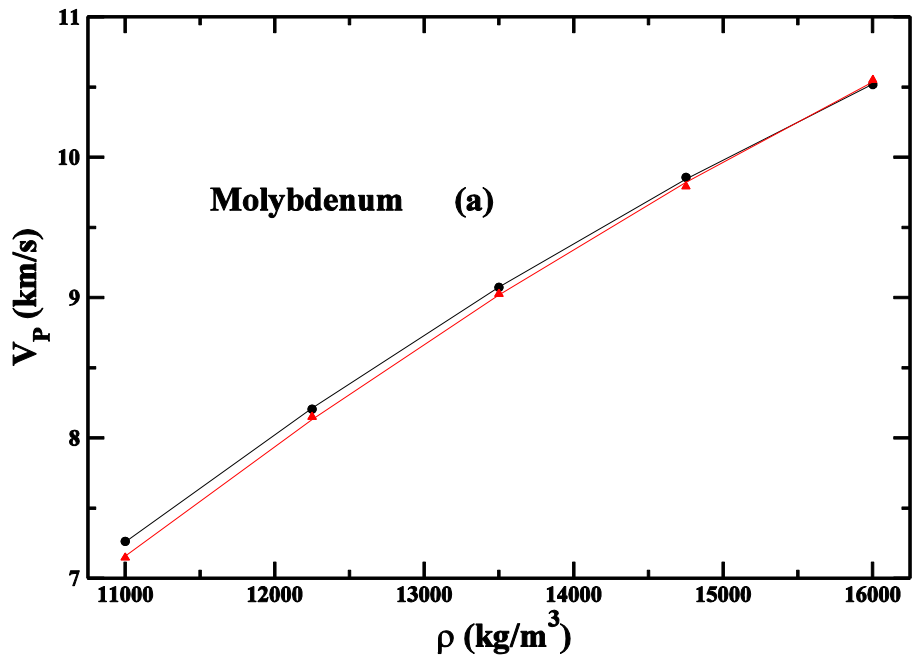

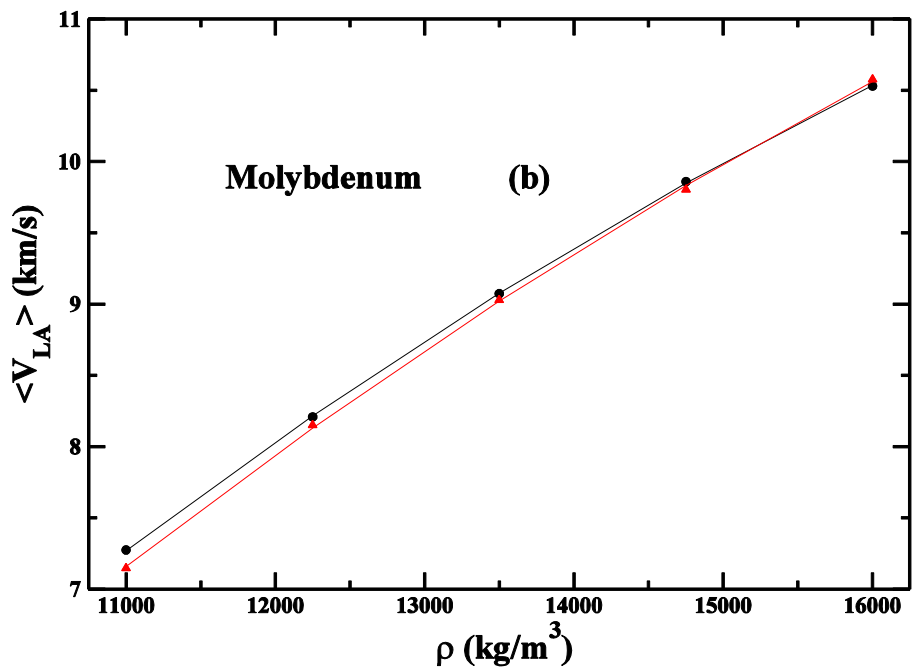

Fig.2: [Color online] (a) P-wave speed ($V_P$) and (b) direction-averaged longitudinal acoustic speed ($<V_{LA}>$) against density ($\rho$) for molybdenum at 300 K (filled circle) and 1500K (filled triangle). Continuous lines are guides for the eye.

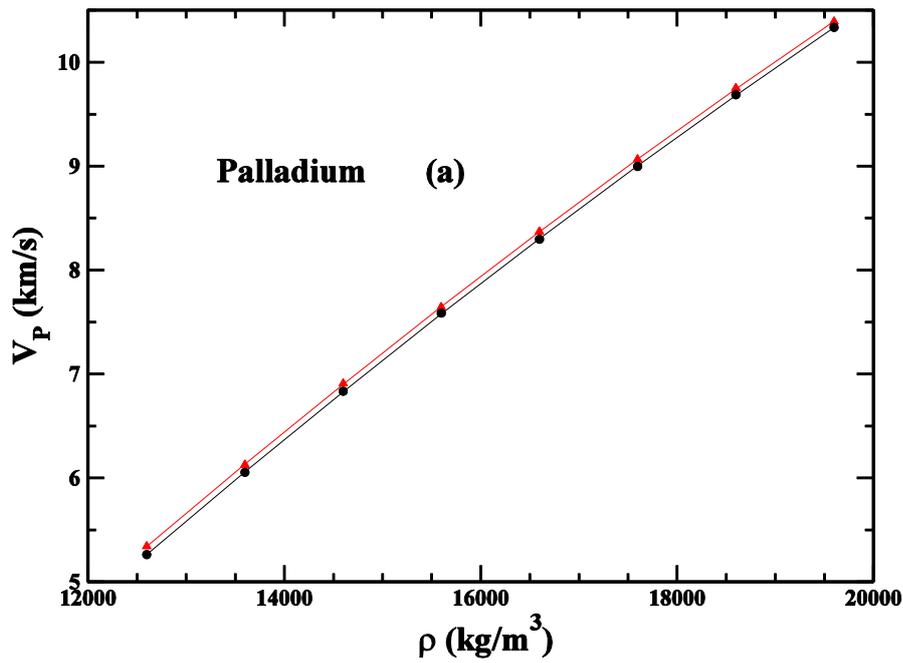

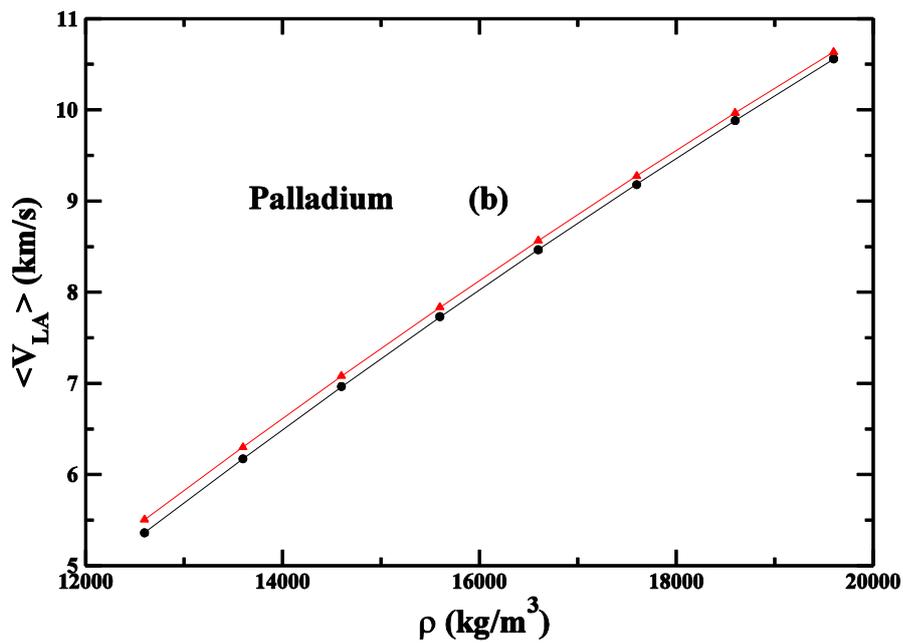

Fig.3: [Color online] (a) P-wave speed ($V_P$) and (b) direction-averaged longitudinal acoustic speed ($<V_{LA}>$) against density (ρ) for palladium at 300 K (filled circle) and 1500K (filled triangle). Continuous lines are guides for the eye.

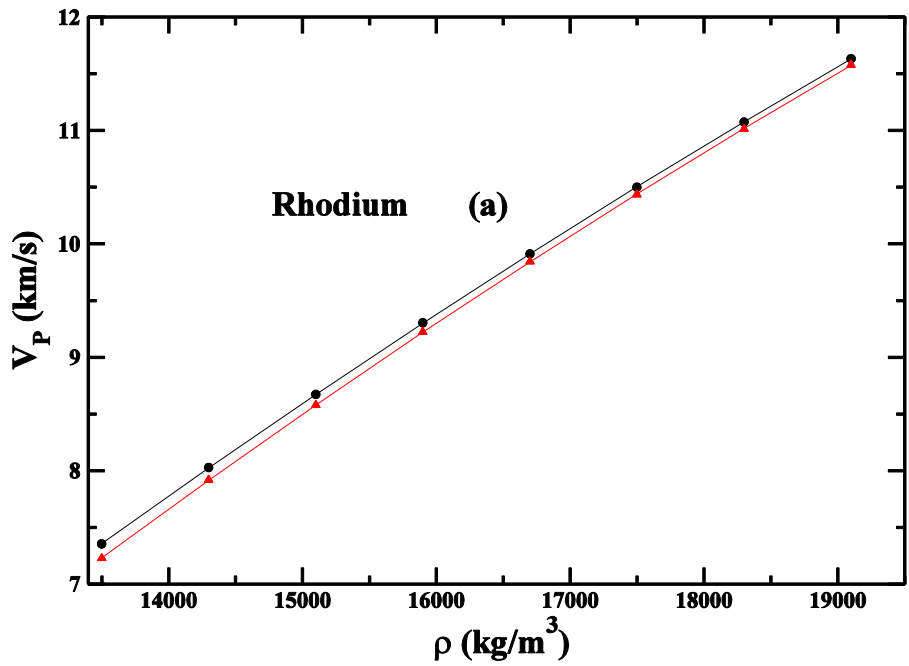

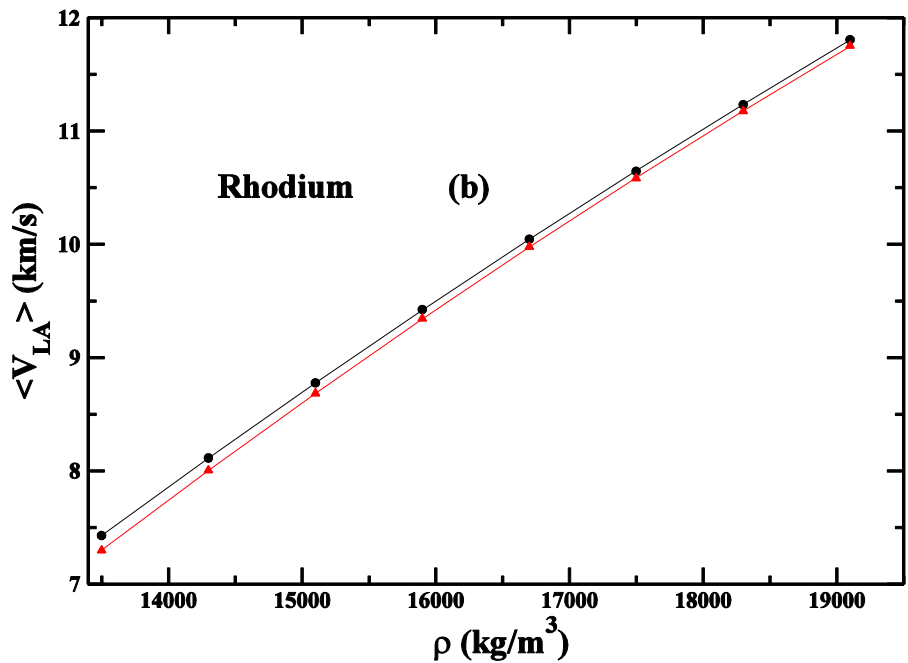

Fig.4: [Color online] (a) P-wave speed ($V_P$) and (b) direction-averaged longitudinal acoustic speed ($<V_{LA}>$) against density (ρ) for rhodium at 300 K (filled circle) and 1500K (filled triangle). Continuous lines are guides for the eye.

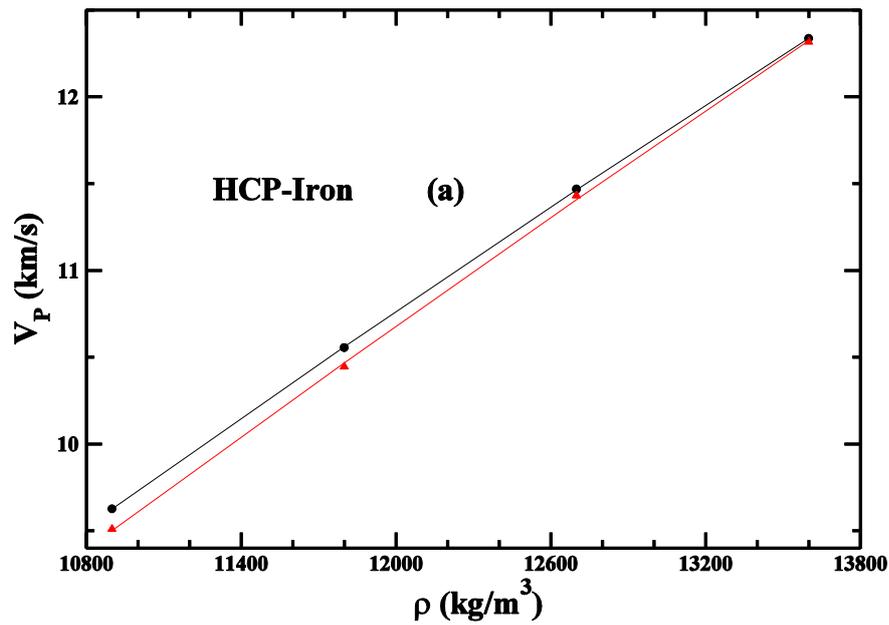

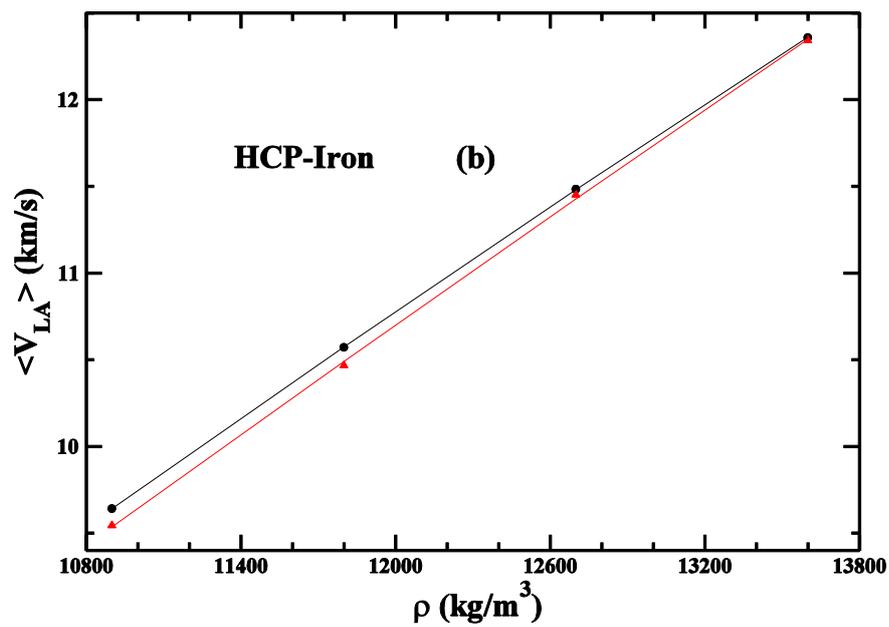

Fig.5: [Color online] (a) P-wave speed ($V_P$) and (b) direction-averaged longitudinal acoustic speed ($<V_{LA}>$) against density ($\rho$) for hcp-iron at 300 K (filled circle) and 1500K (filled triangle). Continuous lines are guides for the eye.

The magnitude of the effect of temperature on the speed versus density plot is generally along intuitively expected lines. A given change of temperature induces a progressively diminishing effect on the change of elastic wave speed as the density increases. Increasing density corresponds to increasing pressure and that reduces lattice vibration and the latter is responsible for a diminishing impact of a change in temperature. However, the anomalous case of molybdenum cannot be explained by this argument. The phase stability of molybdenum at higher pressures and temperatures has been a topic of some debate [52-54] and it is not clear if that has any bearing on our findings. Except for hcp-iron, we have not found any data, computational or experimental, comparable to what are presented in figures 1 through 5. Hence these (especially figs. 1 through 4) should be considered to be predictions that can be verified in laboratory experiments.

To compare the relative merits of the two competing versions of the Birch's law that we set out to test we need to answer the following question: which of the two proposed pairs, (V-$\rho$) or (V$\rho^{1/3}$-$\rho$), has a more accurate linear dependence? One way is to simply visually inspect the two plots – such as shown in figure 6 for the case of platinum at 300K. In this particular case it is obvious that the (V$\rho^{1/3}$-$\rho$) plot is clearly superior when it comes to linearity. However, in general such visual determination may not be possible and it is necessary to define a metric that can answer the question posed above in a quantitative manner.

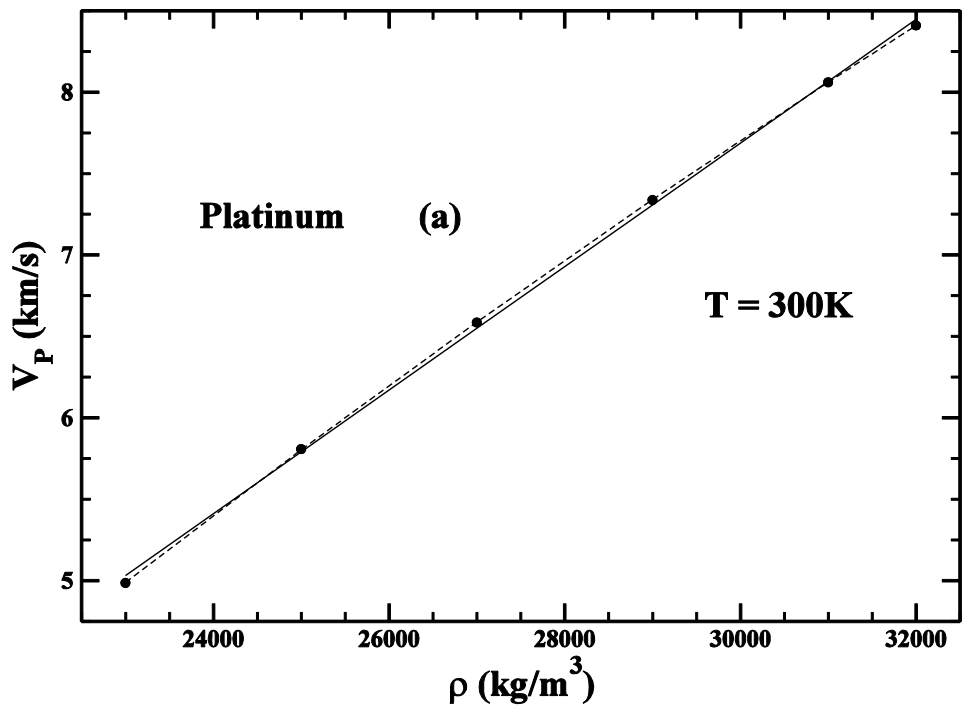

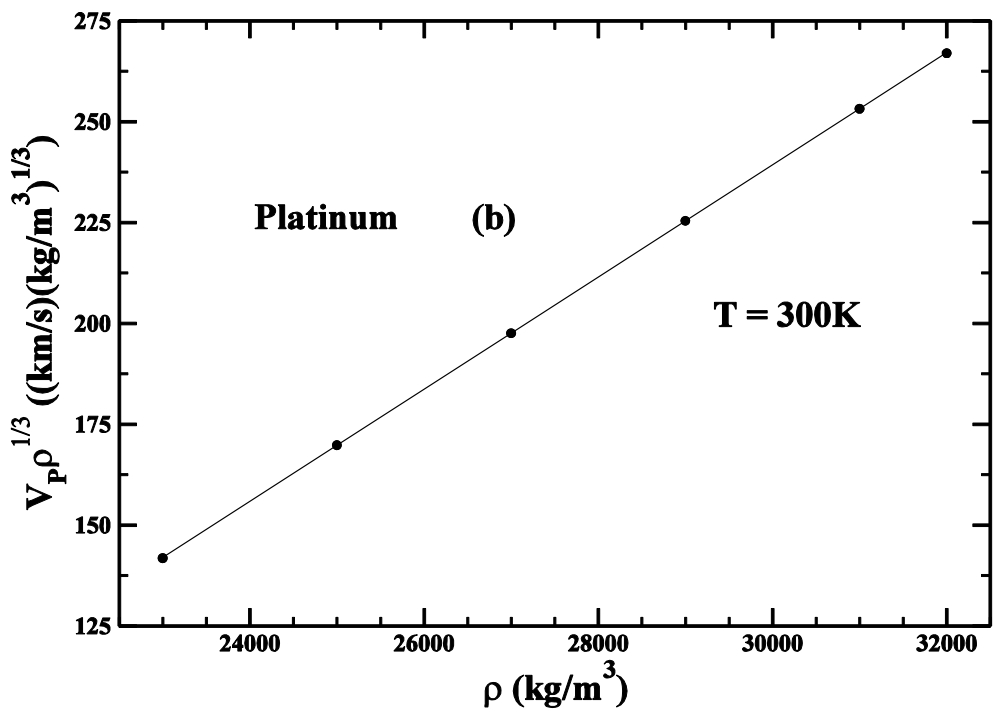

Fig.6: Platinum at 300K: Plot of (a) P-wave speed ($V_P$) and (b) $V_P$ multiplied by $\rho^{1/3}$ vs. $\rho$, where $\rho$ denotes the density. Filled symbols are the data points. Continuous lines are the best fit straight lines. Dashed line in (a) is the best fit quadratic curve.

The metrics we use for this purpose are the maximal and the average fractional deviation of the data points from the best fit straight line --- the data points being the elements of the (V-$\rho$) plot or the (V$\rho^{1/3}$-$\rho$) plot . Here V denotes the elastic wave speed ($V_P$ or $<V_{LA}>$) , $\rho$ denotes the density and fractional deviation is the absolute value of the difference between the observed and the predicted y-coordinates normalized by the difference between the maximum and the minimum values of the y-coordinate with respect to all the data points in that particular graph. The act of using the *fractional difference* (rather than the absolute value of the mismatch) creates a measure that neutralises the difference of dimension between the y-coordinates in the two plots. The idea behind calculating both the maximum and the mean values is to exclude the possibility of drawing a false conclusion due to the existence of any outlier data points. In either case a smaller value of this metric for a given dataset implies a closer agreement with a linear fit.

The values of these metrics are shown in the tables II and III. Inspection of the data in tables II and III shows that: (i) the conclusions drawn regarding the relative superiority of the two alternative versions of the Birch's law does not depend on whether we use the maximum fractional deviation or the mean fractional deviation . This is a reflection of the overall quality and consistency of the data being analysed. (ii) The modified form of the Birch's law consistently gives a more accurate linear fit than the original version in the range of temperatures studied (i.e. up to 1500K) – often by a large margin. This reinforces our conclusions from the previous study at zero temperature [29] -- although the present study seems to suggest that the extent of superiority degrades somewhat with increasing temperature. This issue will be discussed further later.

Table II: Maximum fractional deviation for various datasets

| Element | Temperature in K | Maximum fractional deviation | | | |
|---|---|---|---|---|---|
| | | $<V_{LA}>$ | | $V_P$ | |
| | | $V - \rho$ | $V\rho^{\frac{1}{3}} - \rho$ | $V - \rho$ | $V\rho^{\frac{1}{3}} - \rho$ |
| Pt | 300 | $1.2 \times 10^{-2}$ | $6.9 \times 10^{-4}$ | $1.3 \times 10^{-2}$ | $8.7 \times 10^{-4}$ |
| | 900 | $1.1 \times 10^{-2}$ | $6.8 \times 10^{-4}$ | $1.1 \times 10^{-2}$ | $5.4 \times 10^{-4}$ |
| | 1500 | $9.8 \times 10^{-3}$ | $1.6 \times 10^{-3}$ | $9.7 \times 10^{-3}$ | $2.6 \times 10^{-3}$ |
| Mo | 300 | $2.8 \times 10^{-2}$ | $1.4 \times 10^{-2}$ | $3.6 \times 10^{-2}$ | $1.5 \times 10^{-2}$ |
| | 900 | $2.6 \times 10^{-2}$ | $1.2 \times 10^{-2}$ | $2.8 \times 10^{-2}$ | $1.3 \times 10^{-2}$ |
| | 1500 | $2.7 \times 10^{-2}$ | $1.2 \times 10^{-2}$ | $2.8 \times 10^{-2}$ | $1.3 \times 10^{-2}$ |
| Pd | 300 | $1.6 \times 10^{-2}$ | $1.1 \times 10^{-3}$ | $1.6 \times 10^{-2}$ | $1.3 \times 10^{-3}$ |
| | 900 | $1.5 \times 10^{-2}$ | $1.8 \times 10^{-3}$ | $1.7 \times 10^{-2}$ | $1.6 \times 10^{-3}$ |
| | 1500 | $1.4 \times 10^{-2}$ | $2.0 \times 10^{-3}$ | $1.7 \times 10^{-2}$ | $1.9 \times 10^{-3}$ |
| Rh | 300 | $1.6 \times 10^{-2}$ | $2.3 \times 10^{-3}$ | $1.6 \times 10^{-2}$ | $2.6 \times 10^{-3}$ |
| | 900 | $1.7 \times 10^{-2}$ | $3.2 \times 10^{-3}$ | $1.7 \times 10^{-2}$ | $3.4 \times 10^{-3}$ |
| | 1500 | $1.8 \times 10^{-2}$ | $4.0 \times 10^{-3}$ | $1.8 \times 10^{-2}$ | $3.9 \times 10^{-3}$ |
| hcp-Fe | 300 | $6.1 \times 10^{-3}$ | $1.6 \times 10^{-3}$ | $7.3 \times 10^{-3}$ | $1.5 \times 10^{-3}$ |
| | 1500 | $1.1 \times 10^{-2}$ | $8.8 \times 10^{-3}$ | $1.2 \times 10^{-2}$ | $7.5 \times 10^{-3}$ |

Table III: Mean fractional deviation for various datasets

| Element | Temperature in K | Mean fractional deviation | | | |
|---|---|---|---|---|---|
| | | $<V_{LA}>$ | | $V_P$ | |
| | | $V - \rho$ | $V\rho^{\frac{1}{3}} - \rho$ | $V - \rho$ | $V\rho^{\frac{1}{3}} - \rho$ |
| Pt | 300 | $8.0 \times 10^{-3}$ | $3.8 \times 10^{-4}$ | $8.4 \times 10^{-3}$ | $5.7 \times 10^{-4}$ |
| | 900 | $7.3 \times 10^{-3}$ | $4.1 \times 10^{-4}$ | $7.3 \times 10^{-3}$ | $3.6 \times 10^{-4}$ |
| | 1500 | $6.6 \times 10^{-3}$ | $8.9 \times 10^{-4}$ | $5.9 \times 10^{-3}$ | $1.5 \times 10^{-3}$ |
| Mo | 300 | $2.1 \times 10^{-2}$ | $9.9 \times 10^{-3}$ | $2.3 \times 10^{-2}$ | $1.1 \times 10^{-2}$ |
| | 900 | $2.0 \times 10^{-2}$ | $8.7 \times 10^{-3}$ | $2.1 \times 10^{-2}$ | $9.6 \times 10^{-3}$ |
| | 1500 | $1.9 \times 10^{-2}$ | $8.0 \times 10^{-3}$ | $2.0 \times 10^{-2}$ | $8.6 \times 10^{-3}$ |
| Pd | 300 | $8.8 \times 10^{-3}$ | $5.8 \times 10^{-4}$ | $9.3 \times 10^{-3}$ | $7.4 \times 10^{-4}$ |
| | 900 | $8.6 \times 10^{-3}$ | $7.8 \times 10^{-4}$ | $9.3 \times 10^{-3}$ | $7.3 \times 10^{-4}$ |
| | 1500 | $8.5 \times 10^{-3}$ | $8.3 \times 10^{-4}$ | $9.3 \times 10^{-3}$ | $9.4 \times 10^{-4}$ |
| Rh | 300 | $8.8 \times 10^{-3}$ | $1.4 \times 10^{-3}$ | $8.7 \times 10^{-3}$ | $1.4 \times 10^{-3}$ |
| | 900 | $9.5 \times 10^{-3}$ | $1.9 \times 10^{-3}$ | $9.3 \times 10^{-3}$ | $1.8 \times 10^{-3}$ |
| | 1500 | $1.0 \times 10^{-2}$ | $2.4 \times 10^{-3}$ | $9.8 \times 10^{-3}$ | $2.2 \times 10^{-3}$ |
| hcp-Fe | 300 | $5.0 \times 10^{-3}$ | $8.0 \times 10^{-4}$ | $5.7 \times 10^{-3}$ | $9.7 \times 10^{-4}$ |
| | 1500 | $5.6 \times 10^{-3}$ | $4.4 \times 10^{-3}$ | $6.2 \times 10^{-3}$ | $4.3 \times 10^{-3}$ |

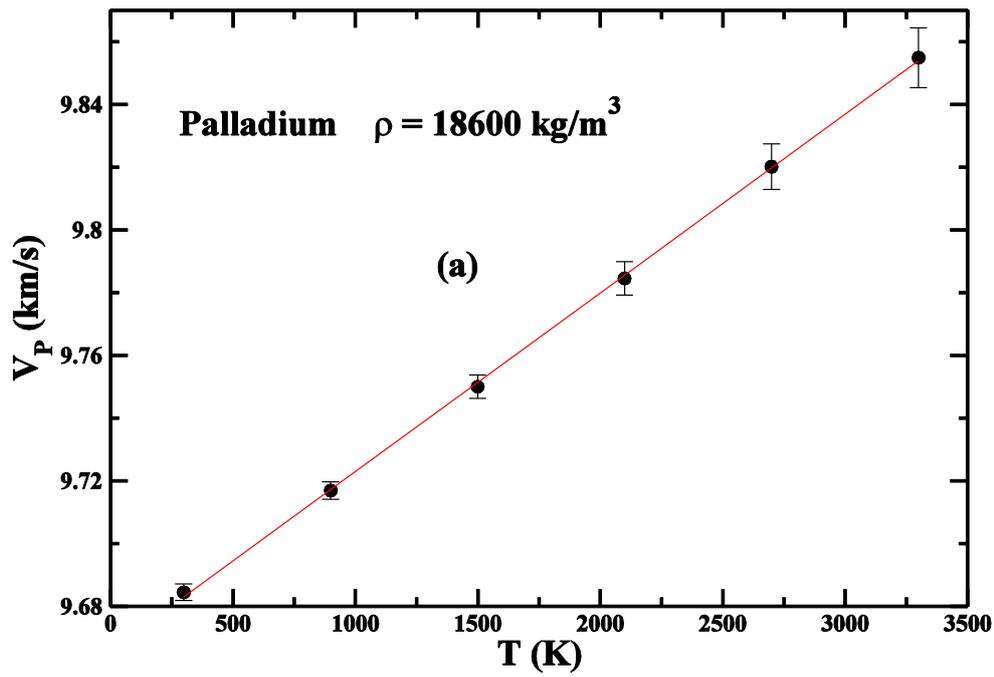

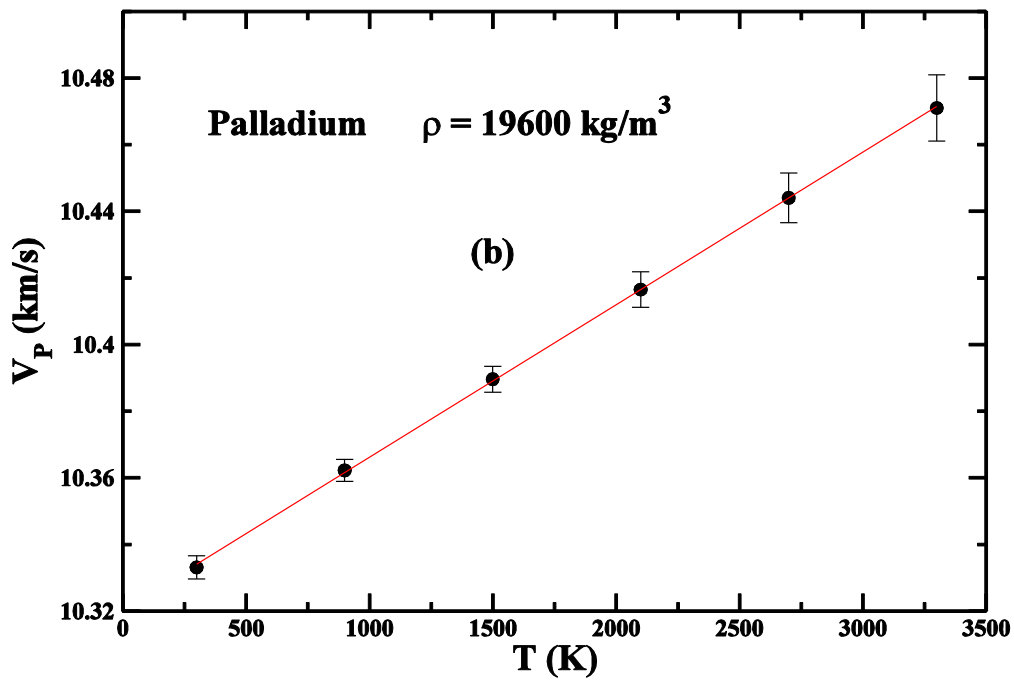

Fig.7: (Color online) P-wave speed ($V_P$) as a function of temperature for palladium at two high values of the density. (a) $\rho$ = 18600 kg/m$^3$ and (b) $\rho$ = 19600 kg/m$^3$. The continuous line is the best linear fit to the data points.

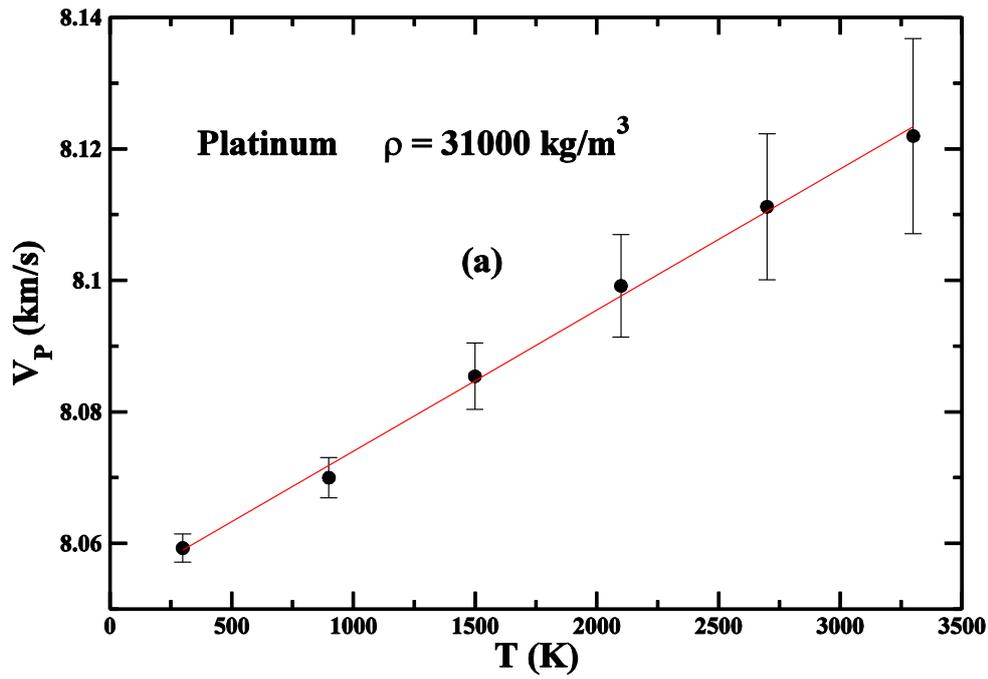

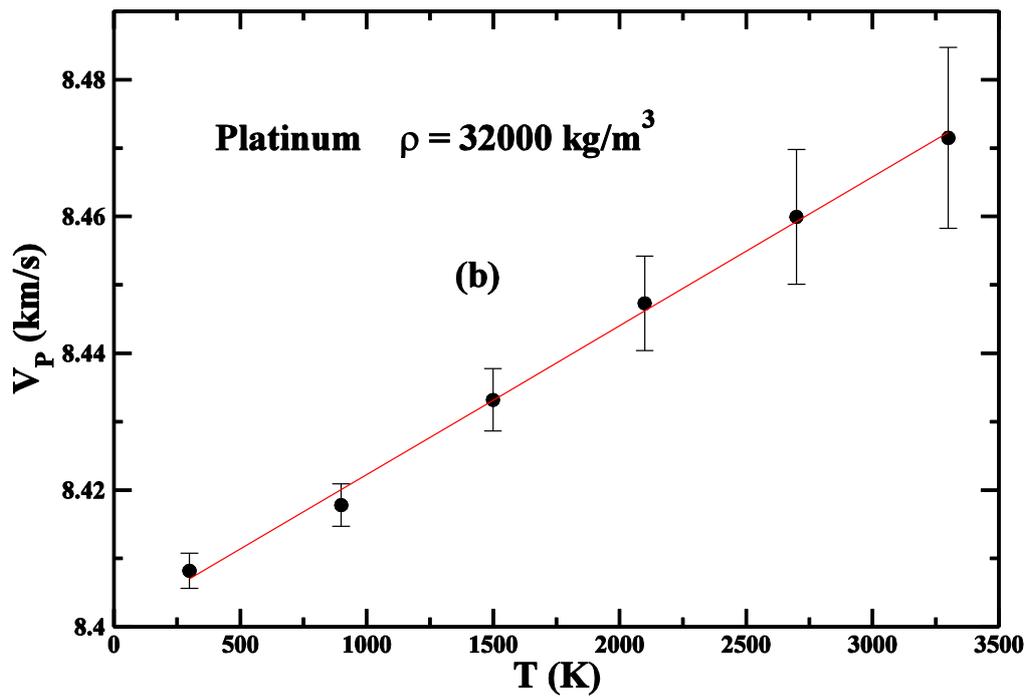

Fig.8: (Color online)   P-wave speed ($V_P$) as a function of temperature for platinum at two high values of the density. (a) $\rho$ = 31000 kg/m³  and (b) $\rho$ = 32000 kg/m³ . The continuous  line is the best linear  fit to the data points.

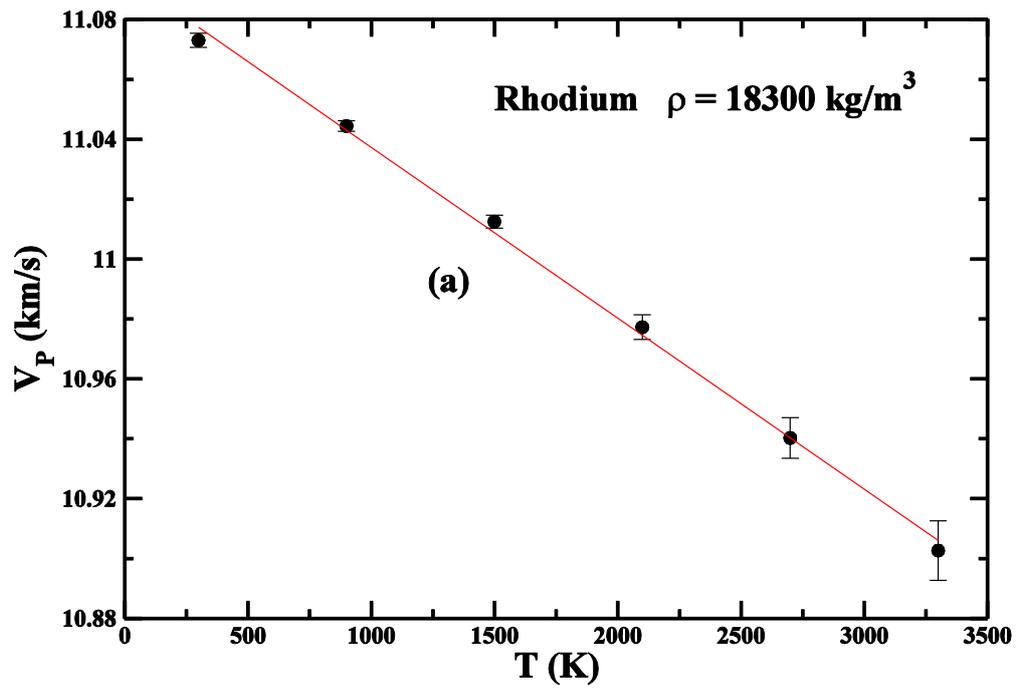

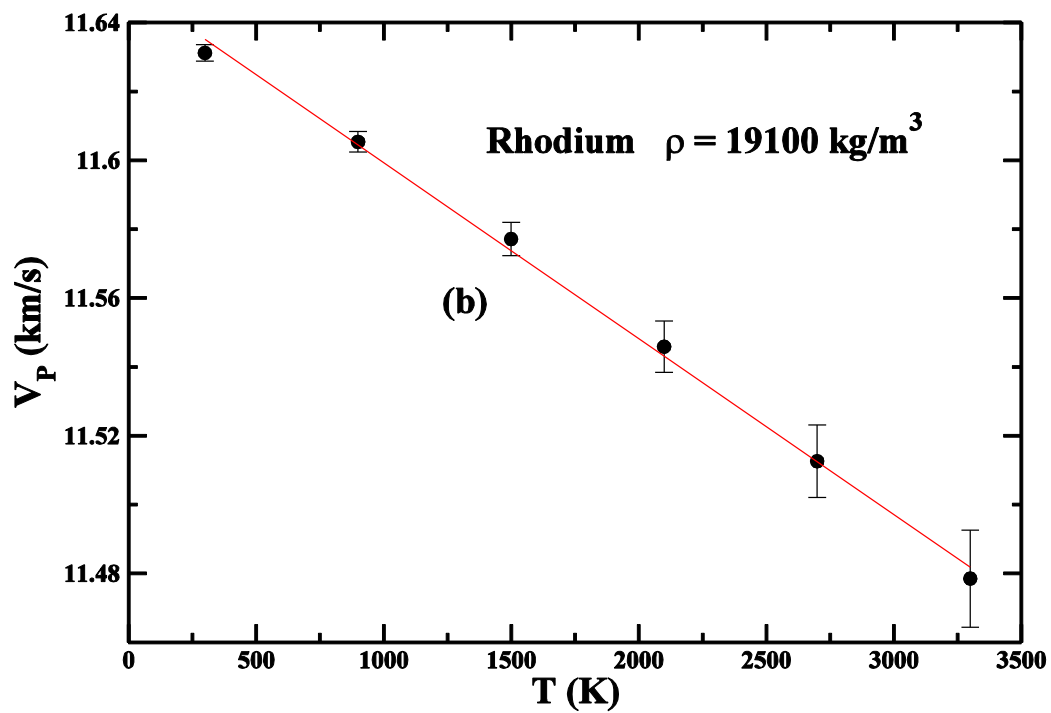

Fig. 9: (Color online)  P-wave speed ($V_P$) as a function of temperature for rhodium at two high values of the density. (a) $\rho$ = 18300 kg/m³ and (b) $\rho$ = 19100 kg/m³. The continuous line is the best linear fit to the data points.

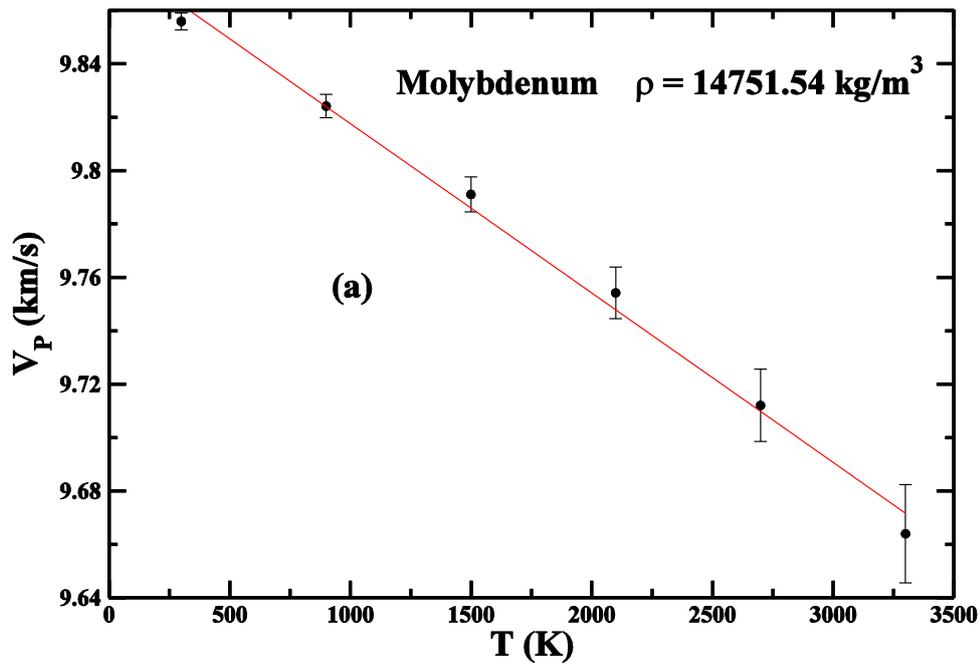

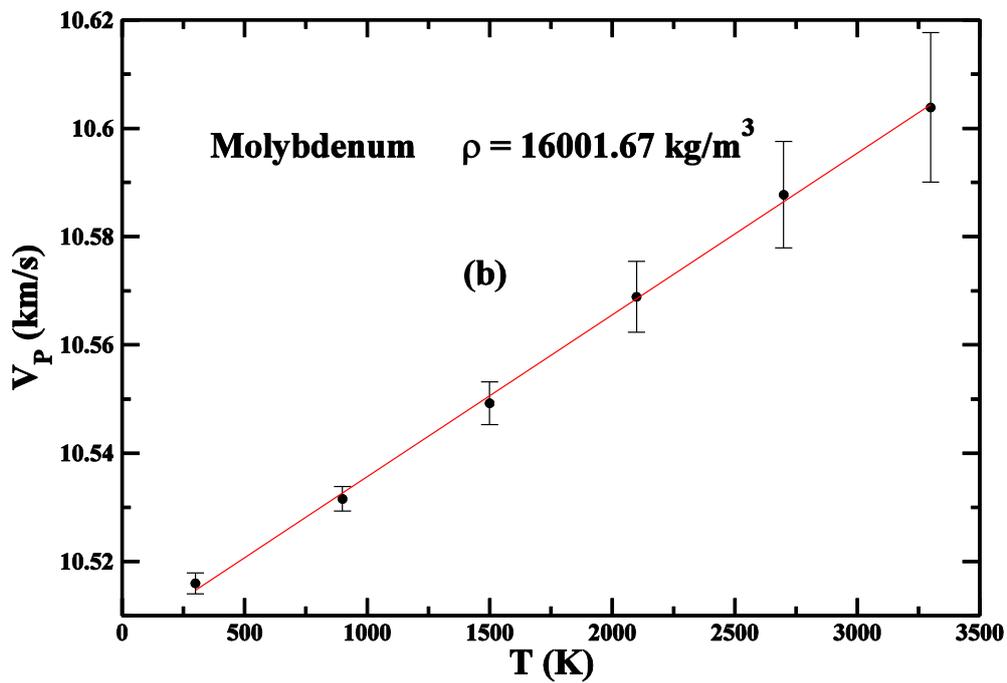

Fig.10: (Color online)  P-wave speed ($V_P$) as a function of temperature for molybdenum at two high values of the density. (a) $\rho$ = 14752 kg /m³ and (b) $\rho$ = 16002 kg/m³ . The continuous line is the best linear fit to the data points.

Finally, we examine an aspect of a recent work [2] on hcp-iron in which the P-wave speed was modelled to be a linear function of temperature at a given density. *This constitutes a de facto extension of the Birch's law* and is of immediate importance to studies connecting laboratory experiments with the question of determining the composition of the inner core of the Earth. We have carried out a test of the genericity of the validity of this idea but only for the cubic materials used earlier in this paper and only at the highest densities – since, at the high temperatures involved, the quasi-harmonic approximation is likely to be valid only when the density is sufficiently high. We show the plots of the P-wave speed against temperature in figures 7 through 10. In each case we also show the best linear fits and the error bars for the computed speeds. The first point to be noted is that the variation with temperature is rather weak in all the cases. This supports the usually held belief that the elastic speed is controlled mostly by the density. The second point, which is the primary purpose of this calculation, is that our data is indeed consistent with the assumption of linearity – although our error bars are somewhat on the larger side.

It may also be noted that the data for molybdenum in fig. 10 agrees with that presented in figure 2 where it was found that the variation of the P-wave speed with temperature changes sign at a critical density. The two values of density used in the figure 10 are on the two sides of this critical density. This, by itself, is no violation of the form of dependence of the P-wave speed on density and temperature suggested in [2] which, in fact, explicitly allows for the possibility of the kind that molybdenum presents.

## IV. DISCUSSION

The primary goal of the present work has been to test the continued applicability of our proposed modification to the Birch's law at higher temperatures. Our conclusion is that, while the conventional form of the Birch's law works fairly well, our alternative version continues to do better (and often significantly better) in terms of the relevant metric. However, the trend in the value of this metric also seems to suggest that this relative superiority suffers an erosion with increase in temperature.

This apparent degradation could be due to a combination of the following reasons: (i) increasing levels of numerical noise as the temperature increases, (ii) error in the calculation of optimal lattice parameters when the reference state is not determined uniquely by the density (this is also influenced by the first cause) -- as in the case of hcp-iron and, of course, (iii) a genuine decline in the ability of the modified Birch's law to describe the phenomenology even in the limit when the first two potential sources of error are entirely absent [Please note that our metric does not differentiate between systematic and statistical sources of error in the value of the elastic wave speed]. From our estimates of the statistical error bars in the calculation of the P- wave velocity at various temperatures we know that indeed the first cause is present. This can be seen quite vividly in the figures 7 through 10 -- as well as in table IV where we present the maximum (with respect to all the densities studied for a particular material) value of the estimate of statistical error in the calculation of $V_P$ for various materials at various temperatures.

Table IV: Maximal estimated statistical error in the computed P-wave speed ($V_P$) for various materials and at various temperatures.

| Material | Statistical Error in $V_P$ (m/s) | | |
| --- | --- | --- | --- |
|  | 300K | 900K | 1500K |
| Molybdenum | 3.2 | 4.9 | 13.3 |
| Palladium | 4.1 | 11.2 | 23.6 |
| Platinum | 2.7 | 3.9 | 7.8 |
| Rhodium | 2.4 | 4.0 | 10.4 |
| Hcp-iron | 7.0 |  | 28.4 |

Without a suppression of this source of error it is not possible to decide conclusively whether the apparent erosion in the superiority of the modified Birch's law is real or is merely caused by the growing temperature induced noise. Moreover, when the specification of the reference state for a given density also requires a (temperature-dependent) optimization of the lattice parameters (as in the case of hcp-iron) error in that calculation also makes an additional contribution to our metric.

We have also analysed each of the four terms in eqn. (3) separately in order to assess their relative contributions to the overall statistical error. For this purpose we performed a cubic fit for each term (as a function of θ) and then calculated the root-mean-square deviation around it. We found that the noise contributed by the phonon term ($F_{ph}$) in the eqn. (3) always increases rather rapidly with temperature. Beyond that the pattern of relative contribution to the noise from the various terms in eqn.(3) is not particularly systematic – at least up to 1500K. Our conclusion on the basis of these observations is that the resolutions for the **q**-grid (and of the **k**-grid to the extent necessary) of the DFT calculations need to be increased significantly in general in order to reduce numerical noises to a level that will allow us to decide more conclusively whether the apparent gradual erosion in the superiority of the modified Birch's law with increasing temperature is genuine. This, of course, implies a much higher computational workload. But without this improvement it is not possible to carry out our tests at higher temperatures more satisfactorily than what we have been able to do. Finally, with increasing temperature the validity of the quasi-harmonic approximation itself will become an issue at some stage.

It may be noted here that the calculation of adiabatic elastic constants implements a direct numerical calculation of the temperature derivatives of the components of the stress tensor. We are not aware of any previous calculation using this procedure. Our experience has been that the plots of the computed thermal stress against temperature (around the target temperature) are smooth enough in all cases to not cause any difficulty in the numerical computation of the derivatives. Also, as mentioned before, the procedure to calculate the isothermal elastic constants does not use any fitting formulas for the free energies but places considerable demand on computational resources to achieve accuracy and reliability. The methodology we have used is general and can, in principle, be used in all types of crystals.

Finally, a few comments are in order on some specific aspects of this work. (i) First of these is our choice of materials to study. By the very nature of our calculations a necessary condition for inclusion in this list was that there should be no phase change (structural or magnetic) in the entire range of densities and temperatures studied for that particular material. The list of elemental solids known to be satisfying this criterion is not particularly large and amongst those that do fulfil this requirement not too many are of direct geophysical interest (The cubic materials that we studied belong to this latter category). But we took the broader view that the Birch's law should be treated and tested in the same manner as any other law of physics would be. It would be extraordinary to think that the validity of a law of nature would be restricted only to a set of materials which are of direct relevance in the study of a particular physical problem. A more utilitarian thought behind this is that unless the law is of sufficiently generic validity it is unlikely to be of much value in situations where the composition is not known with certainty -- as in the studies relating to the inner core of the Earth. (ii) The criterion of magnetic phase stability eliminated nickel -- which is an element of obvious geophysical interest. Although, at room temperature, it is believed [75] to be a ferromagnet in the range of densities studied its Curie temperature is in the middle of the range of temperatures we have investigated -- at least for a subset of densities studied. (iii) Iron has been taken to be in the hcp, non-magnetic phase based on the most recent evidence [76] we are familiar with. (iv) Birch's law is known to hold fairly well for the shear wave speed also in some cases. Hence all the calculations and analysis done for P-waves were performed also for the shear wave (S-wave) speed in every case. However, while the error bars in the calculations of the P-wave and the S-wave speeds are comparable the value of $V_S$ itself is typically about half the value of $V_P$ at a given density. Thus the fractional error bars in the values of $V_S$ are typically more than twice those for $V_P$. As mentioned earlier the metric that we have used to discriminate between the two alternative versions of the Birch's law is not particularly useful in this high fractional error bar situation and no definite conclusions can be drawn. Hence we did not report the analysis for $V_S$ here. (v) One of the key results here is that the P-wave speed can go up or down with increasing temperature. It is a material specific property. Unfortunately, at this time, we do not understand the physics behind this. (vi) It would obviously be of great interest to see if the linear variation of elastic wave speed with temperature, for a given density, holds for iron. However, to be meaningful, this calculation has to be done at a large number of values of the temperatures and would thus be correspondingly more resource intensive.

**SUPPLEMENTARY MATERIAL**

Supplementary Material contains sections on: (A) Procedure for calculating elastic wave speeds, (B) Applications of the Density Functional Theory – including the values of the DFT parameter sets and the values of the computed elastic constants, (C) Analysis of errors in the calculation of the elastic wave speed, (D) Comparison of calculated elastic constants with pre-existing data, and (E) Representative free energy (F) vs. distortion (θ) plots for the cubic case.

**ACKNOWLEDGEMENT**

The computational data reported here were generated using the computing clusters provided to the School of Physical Sciences, Jawaharlal Nehru University by the Department of Science and Technology (DST), Government of India under its DST-FIST-I and DST-FIST-II programs . Additional support by the DST, Government of India under the DST-PURSE program in the form of computing hardware is also acknowledged. UCR would like to acknowledge financial support in the form of a research fellowship from the University Grants Commission, India and in the form of a supplementary fellowship from the DST-PURSE program of DST, Government of India.